\documentclass[floatfix, noshowpacs, preprintnumbers, twocolumn, amsmath, amssymb, aps, prb, superscriptaddress]{revtex4-1}

\usepackage{graphicx}
\usepackage{amsmath,soul,tabularx,mathtools}
\usepackage{amssymb}
\usepackage{times}
\usepackage{bbm}
\usepackage{bm}
\usepackage[T1]{fontenc}
\usepackage{bbold}
\usepackage{nccmath}
\usepackage[mathscr]{eucal}
\usepackage{multirow}
\usepackage{colortbl}
\usepackage[utf8]{inputenc}
\usepackage{lmodern}
\usepackage{xcolor}
\usepackage[caption=false]{subfig}
\usepackage[]{algorithm}
\usepackage{algpseudocode}
\usepackage{booktabs}
\usepackage{hyperref}

\DeclarePairedDelimiter\floor{\lfloor}{\rfloor}

\DeclarePairedDelimiter\abs{\lvert}{\rvert}

\newcommand\PBS[1]{\let\temp=\\%
  #1%
  \let\\=\temp
}

\begin{document}

\title{Emergence of biased errors in imperfect photonic circuits}

\author{Fulvio Flamini}
\affiliation{Institut f\"{u}r Theoretische Physik, Universit\"{a}t Innsbruck, Technikerstraße 21a, A-6020 Innsbruck, Austria}

\begin{abstract}
We study the impact of experimental imperfections in integrated photonic circuits. We discuss the emergence of a moderate biased error in path encoding, and investigate its correlation with properties of the optical paths. Our analysis connects and deepens previous studies in this direction, revealing potential issues for high-precision tests and optical implementations of machine learning.
\end{abstract}

\maketitle

\section{Introduction}

Photonic integrated technologies are a key resource for classical and quantum applications\cite{Wang20,Bogaerts20}, also including machine learning and artificial intelligence \cite{Wetzstein20, Shastri21}. Nevertheless, the unavoidable effect of imperfections due to fabrication, calibration and control can severely undermine their performance \cite{Hollis1990, Fang19}. With path encoding, for instance, a systematic deviation in the scattering probabilities can lead to a biased error in (un)supervised learning \cite{Shen17, Steinbrecher19, Williamson20, Marquez21, Zhang21, Zhang21ACS, Xu21} or reinforcement learning \cite{Flamini19RL, Saggio21}. While major achievements have recently been reported for photonic circuits \cite{Taballione20, Arrazola21, Perez-Lopez21}, error correction \cite{Bandyopadhyay21} and model training\cite{Hughes18, Zhang19, Zhang21ACS}, the inevitability of such imperfections motivates us to shed more light on the above issue.

In this work, we will consider two canonical unitary decompositions \cite{Reck94, Clements16} and discuss potential vulnerabilities in realistic settings. In principle, these analyses can be readily extended also to other optical architectures. We reveal and quantify a mild systematic bias in their operation, which originates from a biased error in the matrix elements of the unitary transformation. We also show how this feature connects to previously known results, and consider a graph-based approach to tackle similar analyses on noise spreading in an optical mesh.
It is not the aim of this work to propose a unique strategy to mitigate these effects, as done very nicely in related works for similar problems \cite{Russell17, Burgwal17, Pai19optimize, Bandyopadhyay21}. We aim to understand these issues and bring them closer to the attention in the field, since practical countermeasures will mostly depend on the actual application and physical constraints.


\section{Framework}

\label{sec:framework}

In this section, we will first review related results in the context of circuit optimization and characterization,
to highlight the main connections and novel contributions. Then, we will introduce the main quantities and models that will be used in our analyses.


\subsection{Related works}

The influence of imperfections in photonic circuits has already been investigated in various works \cite{Clements16, Burgwal17, Flamini17, Pai19optimize}. Here we deepen these studies with a focus on aspects that, to the best of our knowledge, have so far eluded a detailed investigation. Specifically, we study how a biased error can emerge in a noisy circuit, addressing precise, practical questions that concern future applications. 

In realistic settings, it is paramount to improve the operation of imperfect components. Pioneering work in this scope was done by Miller, who proposed solutions for automatic alignment, configuration and compensation \cite{Miller15}. Recent works have focused on various other approaches, including numerical techniques \cite{Mower15, Burgwal17, PerezLopez20}, port allocation and compilation \cite{Kumar21}, and architecture refinements \cite{Burgwal17, Pai19, Fldzhyan20, Bell21}.
Our analysis addresses issues related to error propagation in photonic circuits, so that they can be better addressed by the above strategies.

More closely related to this work, there is a series of studies\cite{Russell17, Burgwal17, Pai19optimize} on the properties of phases and transmissivities in triangular\cite{Reck94}  ($\triangle$) and rectangular\cite{Clements16}  ($\square$) meshes. Ref. \cite{Russell17} presents a strategy to dial up Haar-random unitary transformations by sampling transmissivities and phases from independent probability density functions. Besides its use in practical applications, this study clarifies how these parameters are distributed, which may affect their noise resilience in a non-homogeneous way. Similar considerations have been put forward in Ref. \cite{Burgwal17}, showing that, in larger circuits, the distribution of reflectivities is increasingly skewed towards low values. Given that a non-ideal mesh may not achieve these low values, the overall fidelity of these circuits would be limited. More recently, notions of sensitivities and error tolerances of optical components have been expanded in Ref. \cite{Pai19optimize}, which provides a rich analytic and numerical exploration of these aspects. Our analysis extends the study of unbalanced error tolerance \cite{Pai19optimize} from a different and fine-grained perspective, establishing connections to previous works \cite{Russell17, Burgwal17}. Whenever applicable, we will make these connections explicit. 

The above works focus on theoretical properties of linear optical circuits and error mitigation. However, non-ideal transformations have been object of detailed analyses also in the context of Boson Sampling (see Refs. \cite{Leverrier15, Renema18, Shchesnovich19} and references within), to investigate whether imperfect optical devices preserve classical computational hardness. In this case, the figures of merit are usually global properties of the output probability distribution (e.g. a distance from the ideal distribution). Conversely, the present analysis abstracts from Boson Sampling and targets individual components of the distributions. We will return to this point in Sec. \ref{sec:noisy_multiphoton} and \ref{sec_app:paths}. 

In Sec. \ref{sec:graphs}, we start by introducing the graph-based approach that underlies many of our discussions. In Sec. \ref{sec:uniform_noise} we discuss the core of this work, namely the emergence of systematic deviations in optical meshes subject to uniformly distributed noise. Sec. \ref{sec:calibration} addresses similar considerations for the characterization of reconfigurable circuits. In the Appendix we consolidate the above ideas by proposing more detailed analyses.


\subsection{Universal photonic circuits}

\label{sec:modeling}

\begin{figure}[t]
\includegraphics[width=0.78\linewidth]{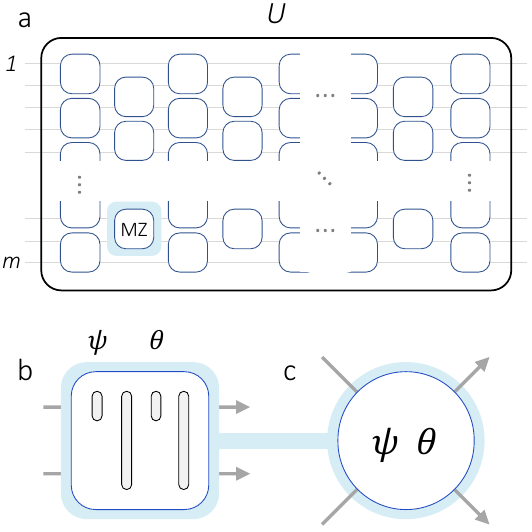}
\caption{\textbf{Circuit layout.} a) Any unitary transformation $U$ can be implemented with a regular mesh of Mach-Zehnder interferometers (MZ), which apply a transformation on two adjacent modes according to well-known decompositions \cite{Reck94, Clements16}.  b) An ideal MZ consists of two 50:50 beamsplitters interleaved with two phase shifters ($\psi$, $\theta$). c) The action of a MZ can be modelled as a node in a directed graph, where information and errors propagate from left to right.
}
\label{Fig1}
\end{figure}

Any $m$-dimensional unitary transformation $U$ can be implemented by an integrated photonic circuit using only meshes of beamsplitters and phase shifters (Fig. \ref{Fig1}a). Here we focus on two canonical architectures, the $\triangle$ \cite{Reck94} and $\square$ \cite{Clements16} meshes, although other approaches exist in the classical  and quantum domains \cite{Bogaerts20, Wang20, Dhand15, deGuise18, Su19}. These architectures consist of optical meshes of Mach-Zehnder interferometers (MZ), arranged in the regular geometry described by their name. An ideal MZ consists of a pair of 50:50 beamsplitters interleaved with a pair of phase shifters $(\psi, \theta)$ (Fig. \ref{Fig1}b), whose action on the two input modes is described by the transformation

\begin{equation}
 U^{MZ} =   
    \begin{bmatrix}
   1 &   i \\    i & 1 
   \end{bmatrix}
     \begin{bmatrix}
   e^{i \psi} &   0 \\    0 &   1
   \end{bmatrix}
     \begin{bmatrix}
   1 &   i \\    i & 1 
   \end{bmatrix}
   \begin{bmatrix}
   e^{i \theta} &   0 \\    0 &   1
   \end{bmatrix}.
   \label{eq:MZ}
\end{equation}

There exist also equivalent formulations \cite{Miller15} where the first phase shifter lays on the output mode or in the lower arm of the MZ. With this notation, when $\psi=0$ the MZ is in the \textit{cross} state (all light is transmitted), while $\psi=\pi$ gives the \textit{bar} state (all light is reflected). Experimental imperfections are usually modelled by adding Gaussian noise to the ideal parameters. For beamsplitters, this corresponds to sampling a parameter around the ideal value of the transmission amplitude ($1/\sqrt{2}$), with a width given by the state of the art. 

Any $m \times m$ $U$ can be decomposed\cite{Reck94, Clements16} in the product of a diagonal matrix $D$ (with complex unit magnitude elements) and $m(m-1)/2$ unitaries $U_k$

\begin{equation}
 U = D \prod\nolimits_k U_k
\end{equation}

\noindent implementing $m$-mode Givens rotations on the planes spanned by two adjacent optical modes 

\begin{equation}
 U_k =
\begin{bmatrix} 
    1       & \hdots    & 0         & 0         & \hdots    & 0\\
    \vdots  & \ddots    & \vdots          & \vdots    &           & \vdots\\
    0       & \hdots    & u_{11}    & u_{12}   & \hdots    & 0\\
    0       & \hdots    & u_{21}    & u_{22}   & \hdots    & 0\\
    \vdots  &           & \vdots    & \vdots    & \ddots    & \vdots\\
    0       & \hdots    & 0         & 0         & \hdots    & 1
    \end{bmatrix},
\end{equation}

\noindent where the $u_{ij}$ ($i,j \in [1,2]$) are the matrix elements of a MZ (see Eq. \ref{eq:MZ}). In the following sections, for many of our analyses we will also represent optical circuits as graphs (see Fig. \ref{Fig1}b,c and the next section).

\begin{figure*}[ht]
\includegraphics[width=0.95\textwidth]{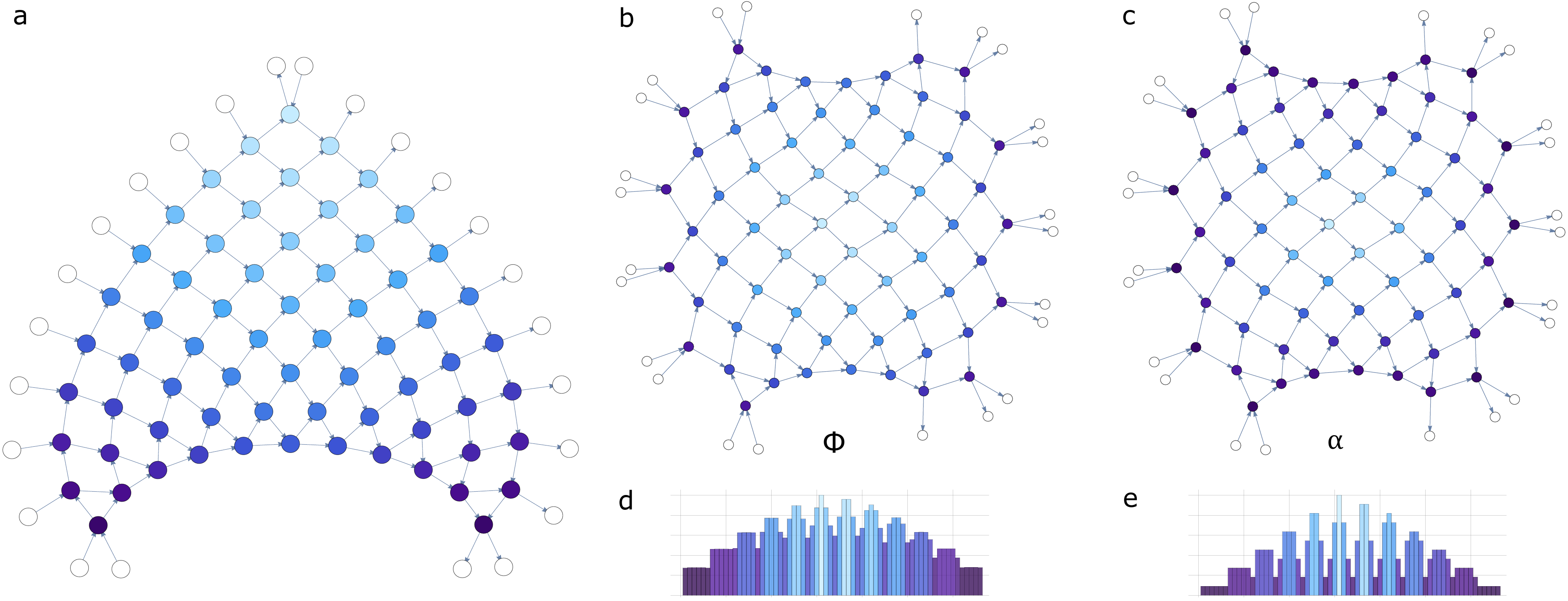}
\caption{\textbf{Photonic circuits as graphs.} $\triangle$ and $\square$ architectures can be represented as planar graphs to study error propagation. Here, white nodes represent $m=12$ input/output modes, colored nodes represent unit cells (two phase shifters each), while edges represent waveguide connections. Different measures of node centrality can be studied and visualized with this approach (see Sec. \ref{sec_app:centrality}). a) $\triangle$ topology. Colors represent the value of the flow $\Phi$ for each node (see Eq. \ref{eq:flow}), normalized by the maximum. Lighter nodes have a higher value, as in panels (d,e). b) $\square$ topology, same color scheme as panel (a). c) Sensitivity measure proposed in Ref. \cite{Pai19}. d,e) Data from panels (b,c) projected onto a bar chart (using the same ordering, layer by layer).}
\label{Fig2}
\end{figure*}


\section{Photonic circuits as graphs}

\label{sec:graphs}

A convenient way to investigate the effect of imperfections in photonic circuits is to represent them as directed graphs. For $\triangle$ and $\square$ topologies, such graphs are also acyclic: light enters from a subset of the $m$ input nodes, propagates from left to right and leaves the graph from the $m$ output nodes. This approach, recently discussed in a different context in other works \cite{Pai19, Chen20}, facilitates the study of error propagation. In this section we make this intuition more concrete. 

A graph representation for the canonical architectures is shown in Fig. \ref{Fig2}a ($\triangle$) and \ref{Fig2}b,c ($\square$). This approach inspires a number of questions about symmetry: are there more influential nodes contributing to error spreading? Similar questions have already drawn large attention in areas related to information dissemination, contagious disease transmission and social influence \cite{Fu15}. In these cases, nodes with a high centrality index (see Sec. \ref{sec_app:centrality}) are known to facilitate spreading, even though this depends on many factors. Motivated by this connection, we propose to leverage ideas from interaction networks to study errors in optical networks.

A notion of sensitivity has been introduced for MZs that depends on their position in the mesh \cite{Pai19}. This quantity is linked to the shape of the distributions induced by the Haar measure over the physical parameters \cite{Russell17, Burgwal17}. Here we show how consistent considerations can be made from a different perspective, by considering the importance of each MZ (node $n$) in the circuit (graph $G$). We investigate two aspects: centrality measures (discussed in Sec. \ref{sec_app:centrality}) and properties related to paths in a graph. The latter approach is convenient to study error propagation in waveguide structures and, assuming a homogeneous level of noise over all optical elements, to estimate the amount accumulated in the circuit. If we focus on a node, it is reasonable to expect that its influence increases the more connected it is to the input and output modes. We can even focus on subsets of input/output modes, or associate different weights with different paths. For instance, we can expect a node to be more influential if paths are on average longer, since losses and noise increase with the number of elements. To make it more quantitative, we introduce the following heuristic figure of merit, which we simply call $flow$, 
 
\begin{equation}
    \Phi^{(k)}_G{(n)} \coloneqq \sqrt{\Phi^{(k)}_{G,in}(n) \; \Phi^{(k)}_{G,out}(n) }
    \label{eq:flow}
\end{equation}

\noindent where

\begin{equation}
    \begin{split}
        \Phi^{(k)}_{G,in}(n) \coloneqq \frac{1}{|I_n|} \sum_{i \in I_n} \langle |\textup{path}|^k \rangle_{ \{\pi_{i,n}\} }   \\
        \Phi^{(k)}_{G,out}(n) \coloneqq \frac{1}{|O_n|} \sum_{j \in O_n} \langle |\textup{path}|^k \rangle_{ \{\pi_{n,j}\} } ,
    \end{split}
\end{equation}

\noindent $I_n$ ($O_n$) is the set of input (output) nodes in $G$ reaching (reachable by) node $n$, $\{\pi_{a,b}\}$ is the set of all paths $\pi_{a,b}$ from node $a$ to $b$, and $|.|$ is their length, for some $k>0$. For $k=1$, the flow is the geometric average of the mean lengths of all paths connecting input and output to node $n$. The intuition is that the higher $\Phi_G(n)$, the more $n$ is likely to be critical in a noisy setting. Results for $k=1$ are shown in Fig. \ref{Fig2} ($k=0.5$ or $k=2$ behave qualitatively similarly). We find that $\Phi_G(n)$ is consistent with the sensitivity index $\alpha_G(n) = |I_n| + |O_n| + m -1$ introduced in Ref. \cite{Pai19}, since they correlate well for both $\triangle$ and $\square$ (Fig. \ref{Fig2}b,c) and since, for a given $m$, they both depend only on $I_n$ and $O_n$. The main difference is that, while $\alpha_G(n)$ has a solid theoretical support that is lacking in the present discussion, $\Phi_G(n)$ can in principle be applied to circuits with arbitrary topologies. Also, while $\alpha_G(n)$ is an effective and compact tool to study Haar-random unitaries, $\Phi_G(n)$ is in principle compatible with applications where unitaries are not drawn according to this measure. It would be exciting to design an estimator that generalizes to arbitrary topologies and that is more grounded in the theory of interferometric design.

\begin{figure}[b]
\includegraphics[width=0.375\textwidth]{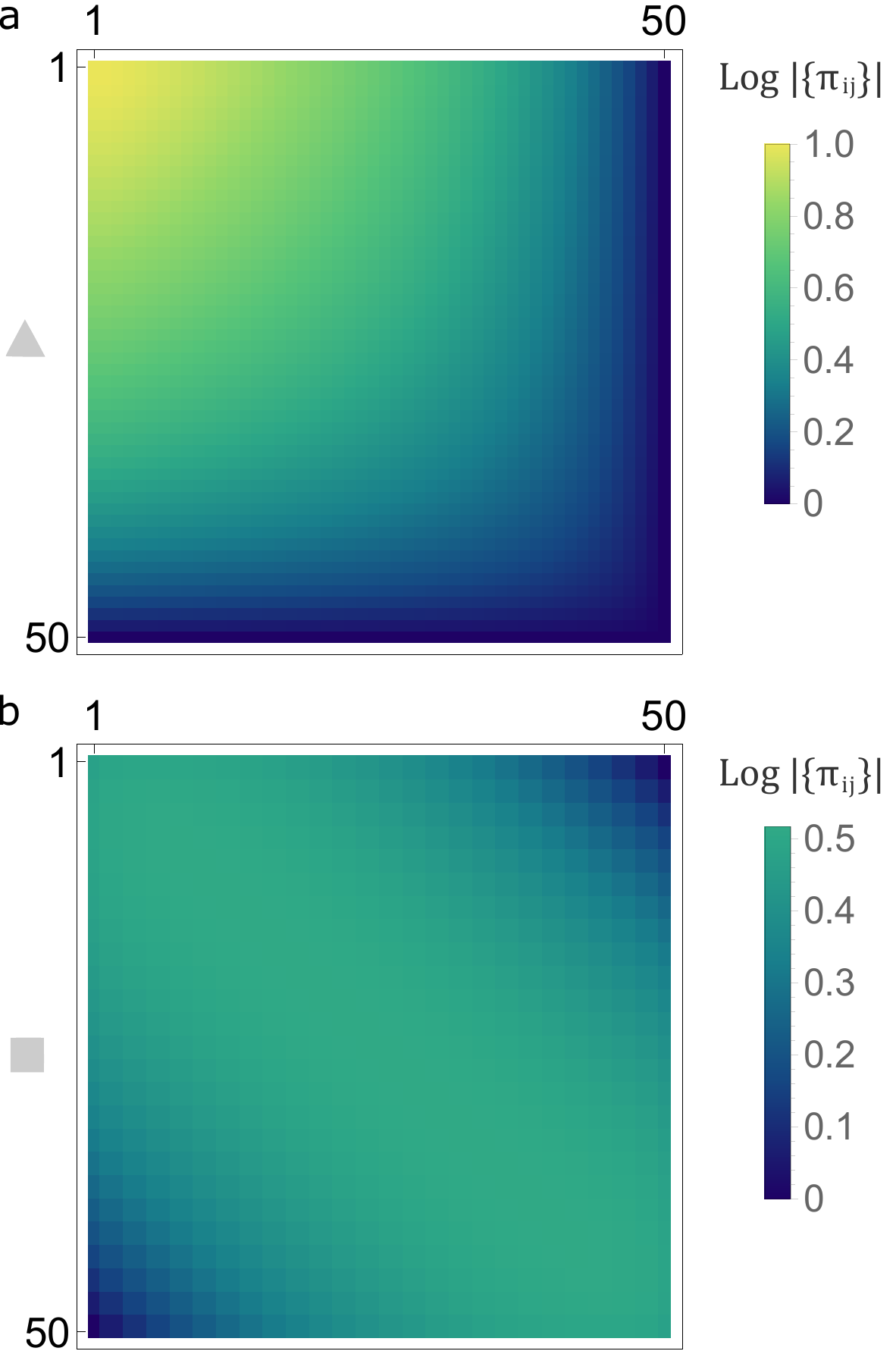}
\caption{\textbf{Optical paths.} Logarithm of the number of paths $\pi_{i,j}$ that connect the input modes ($j$, columns) to the output modes ($i$, rows) of a $\triangle$ (a) or a $\square$ (b) architecture ($m=50$), normalized by the maximum over the two plots (colors can be compared element-wise between the panels). Notice that here each cell relates to an input/output combination, while in Fig. \ref{Fig2}a-c each node of the graph represents a MZ.
}
\label{Fig3}
\end{figure}

The above analysis focused on individual nodes $n$ inside the circuit. When $n$ is an output node, the analysis concerns the connection between optical paths and the matrix elements $u_{ij}$ that describe a scattering process. Specifically, we can count the number of paths $\pi_{i,j}$ that connect any pair ($i,j$) of input and output modes. This information allows to reveal the imbalance in the number of tunable elements that influence all $u_{ij}$, within the same topology and for different topologies (see also Sec. \ref{sec:noisy_multiphoton}). We find that this number is symmetric about the main diagonal ($i=j$), due to the vertical symmetry of the mesh, while $\square$ is almost symmetric about both. As expected, we also find that $u_{ij}$ in $\triangle$ present a strong imbalance w.r.t. paths, while the $\square$ architecture is significantly more balanced. Our aim is then to quantify this intuitive discrepancy, which is one of the reasons why $\square$ is usually considered superior. However, the relative ease of a brute-force approach (convenient to explore arbitrary, small circuits up to $m \sim 30$) comes at the expense of a computationally intensive evaluation, given the exponentially large number of paths. Specific tricks (e.g. leveraging symmetries or propagation in light cones) can only mildly extend this range. Hence, we looked for a more efficient way to estimate the number of paths that contribute to the scattering amplitude $u_{ij}$. We find that an exact, analytical solution can be obtained using a special integer sequence given by the Catalan's trapezoid \cite{Reuveni14}. We discuss its derivation in Sec. \ref{sec_app:paths}, with further considerations on the asymptotics. We numerically verify its prediction by comparing it to the graph-based approach, up to the largest size that could be probed on a workstation. Fig. \ref{Fig3} shows an example for $m=50$ (no new pattern emerged for larger $m$).

\begin{figure}[h]
\includegraphics[width=0.475\textwidth]{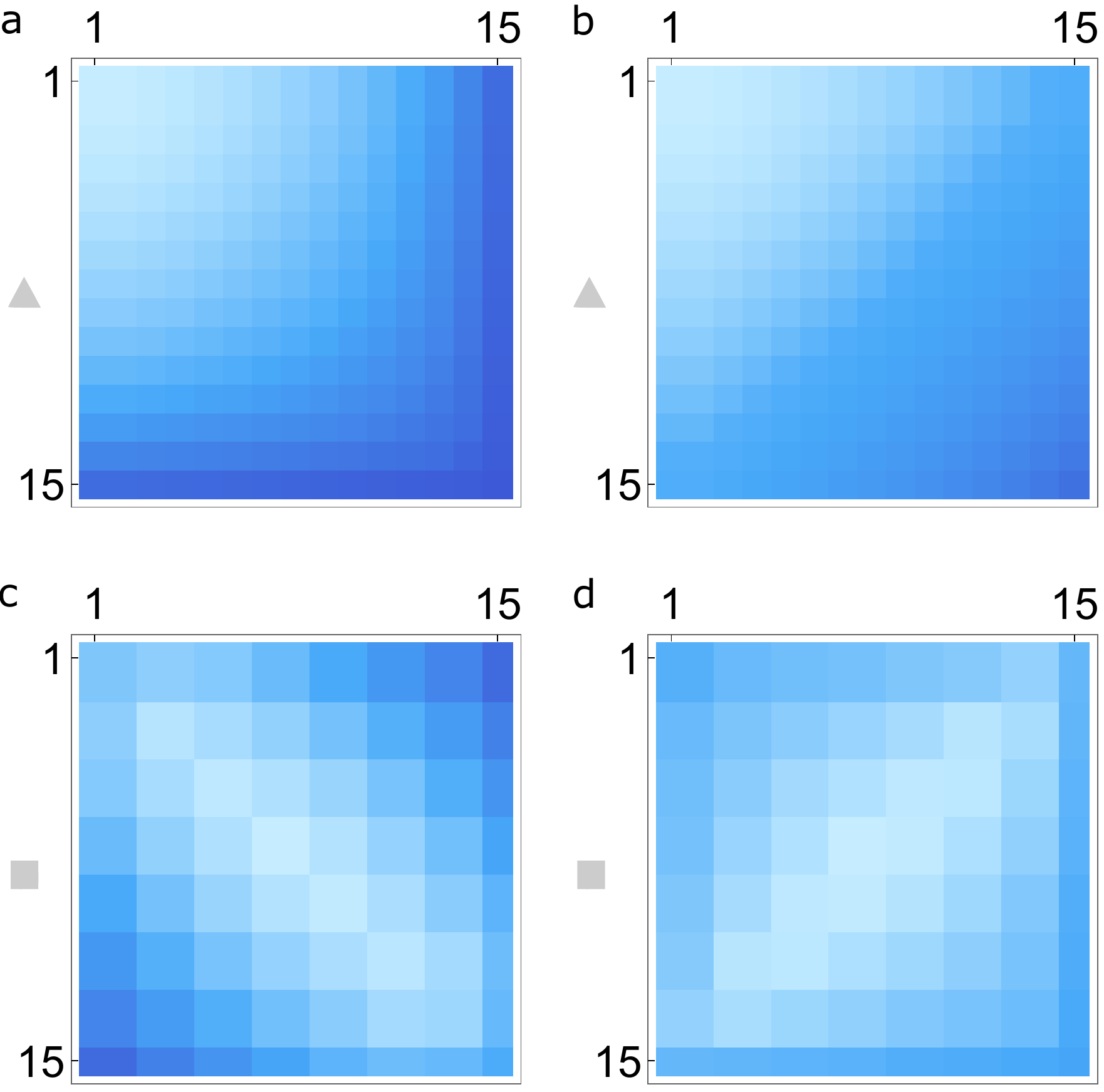}
\caption{\textbf{Path lengths.} Logarithm of the total length (a,c) and average length (b,d) of all paths $\pi_{i,j}$ for $\triangle$ (a,b) and $\square$ (c,d) architectures ($m=15$). Values are normalized w.r.t. the maximum in each plot (the brighter the color, the higher the value). Notice that all panels have the same number ($15 \times 15$) of grid elements: for $\square$ meshes in panels (c,d), modes connected to the same input/output MZs behave in the same way. This analysis complements that reported in Fig. \ref{Fig3}.
}
\label{Fig4}
\end{figure}

In general, the formula presented in Sec. \ref{sec_app:paths} - and estimators analogous to Eq. \ref{eq:flow} - could be used to quantify the importance of a node for a transition, as well as the asymmetry between transitions.
Conversely, the graph-based approach is flexible, can inspire the design of new estimators (see Fig. \ref{Fig4}) and supports arbitrary topologies, for which a closed-form expression is not available. Together, these considerations set the stage to discuss how biased errors emerge in single- and multi-photon dynamics in linear optical circuits.


\section{Uniform noise and nonuniform deviations in a scattering process}

\label{sec:uniform_noise}

Given the difference between $\triangle$ and $\square$ meshes shown in Fig. \ref{Fig3} and \ref{Fig4}, we ask whether this is reflected in the noise sensitivity of the elements $u_{ij}$ of a unitary implemented with these schemes. This analysis aims to reveal a potential bias that has eluded studies based on coarse-grained measures, such as the total variation distance or the fidelity. Indeed, while these measures provide a simple and useful summary of the overall quality of an implementation, they may not reflect the needs of a practical application. For instance, we do not want a classifier with a different accuracy for different objects (or we would like to compensate for this asymmetry).  Also, we expect that high-precision applications, e.g. tests in quantum foundations\cite{Gstir21}, would benefit from a fine-grained understanding of the noise resilience of the scattering matrix.


\subsection{Single-photon evolution}

\label{sec:singlephoton}

\begin{figure*}[h!t]
\includegraphics[width=0.98\textwidth]{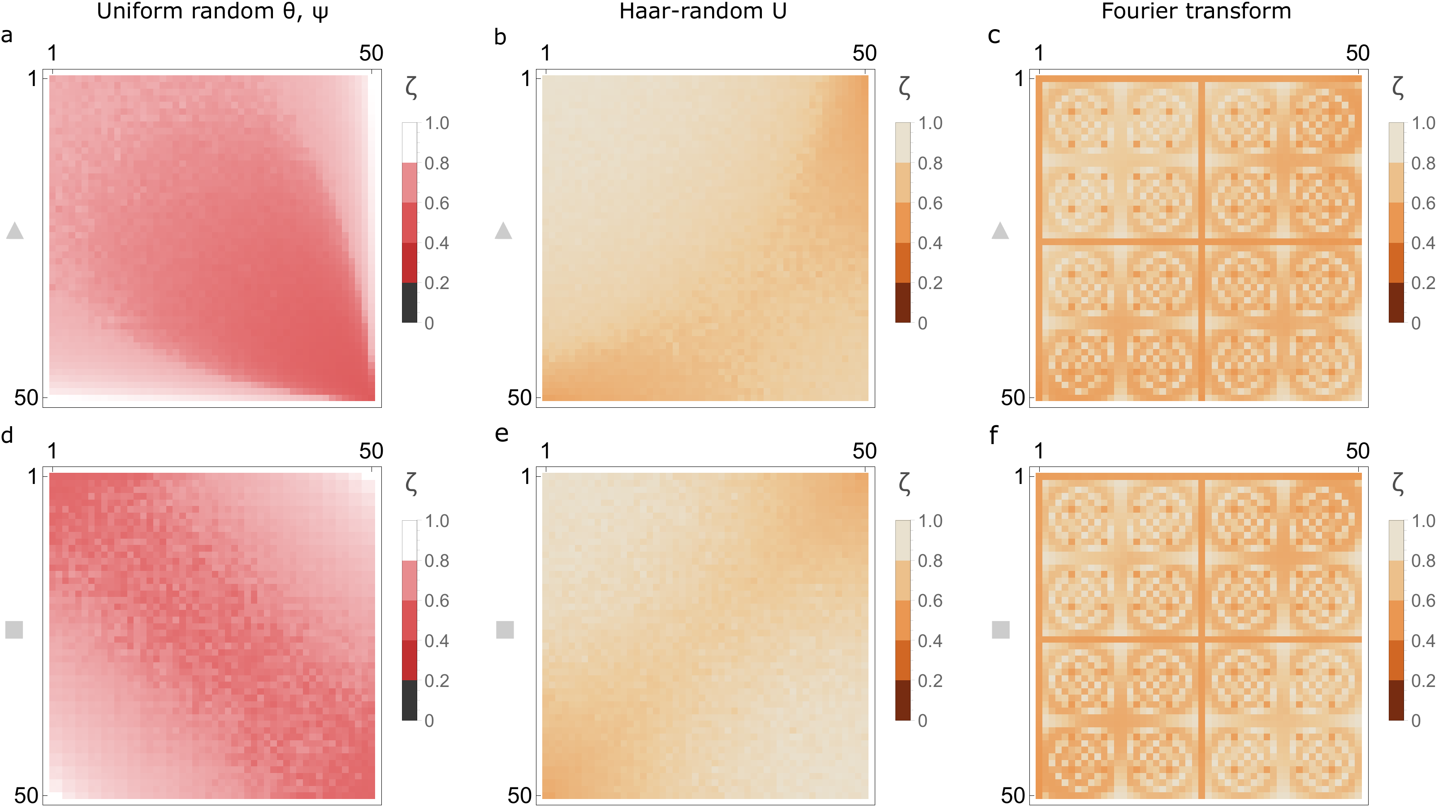}
\caption{\textbf{Sensitivity of the matrix elements.} The noise sensitivity of $u_{ij}$ reflects patterns in the input-output connectivity (see Fig. \ref{Fig3} and \ref{Fig4}), but more rigorous analyses are necessary for a full understanding of this correlation. The overall patterns given by $\zeta_{ij}(U)$ in Eq. \ref{eq:estimator_bias} depend on how parameters are sampled (top row: $\triangle$ meshes; bottom row: $\square$ meshes). For instance, patterns for Haar-random unitaries are due to the distribution of phases in the circuit\cite{Burgwal17}.
We study this sensitivity by numerically generating $2 \times 10^4$ 50-mode unitaries $U$, for $2 \times 10^3$ random initializations of imperfect components each. Tests are carried out using uniformly sampled phases (a, d), Haar-random unitaries (b, e), and the Fourier transform (c, f). Imperfections are modelled with a mild Gaussian noise on the amplitude transmission of the 50:50 beamsplitters ($\mu=2^{-1/2}$, $\sigma=10^{-2}$) and on the phase shifters ($\sigma=10^{-3}$). In all panels, brighter colors correspond to elements that deviate the most from the ideal value.}
\label{Fig5}
\end{figure*}

To investigate this aspect, we quantify the average deviation of the noisy element $\tilde{u}_{ij}$ when Gaussian noise is added in the architecture (i.e. not directly to $u_{ij}$). The research question we address here is whether the sensitivity of $u_{ij}$ correlates with the patterns found in Fig. \ref{Fig3} and \ref{Fig4}. A positive answer would suggest an explanation for this phenomenon, which is the first step to counteract it. To this end, we consider the following quantity

\begin{equation}
    \zeta_{ij}^{(f)}(U) \coloneqq \abs*{ 1-2  \left(1+\abs*{\frac{f(u_{ij})}{f(\tilde{u}_{ij})}} \right)^{-1} },
\end{equation}

\noindent which is nothing more than an adaptation of the interferometric visibility, for some suitable complex function $f$. Choosing $f_1(z)=\textup{Re}(z)$ and $f_2(x)=\textup{Im}(z)$ and considering several random imperfect implementations for each $U$, we can compute the estimator

\begin{equation}
    \zeta_{ij}  \coloneqq \gamma_1 \, \big \langle \zeta_{ij}^{(\textup{Re})}(U) \big \rangle_U + \gamma_2 \, \big \langle \zeta_{ij}^{(\textup{Im})}(U) \big \rangle_U,
    \label{eq:estimator_bias}
\end{equation}

\noindent where $\gamma_{1,2}$ normalize the two terms by their maximum, for both architectures and different unitaries. We consider two ways to sample $U$: according to the Haar measure (uniform over the unitaries, not in the phase space \cite{Burgwal17}) and by uniformly sampling the tunable phases in the circuit (not uniform over the unitaries).
Results reveal a clear \textit{imbalance} in the sensitivity to a \textit{homogeneous} Gaussian noise among the matrix elements of $U$ (see Fig. \ref{Fig5}).

In the case of uniformly sampled phases (Fig. \ref{Fig5}a,d), we find that the sensitivity of $u_{ij}$ (vaguely) negatively correlates with the number of optical paths: the higher the connectivity is (see Fig. \ref{Fig3} and \ref{Fig4}), the lower the deviation tends to be as measured by $\zeta_{ij}$. However, the pattern that emerges in Fig. \ref{Fig5}a suggests that a thorough explanation depends also on other factors. From a practical standpoint, this imbalance prompts us to take it into account when unitaries are not sampled from the Haar measure, e.g. when training machine learning models.

In the case of Haar-random unitaries (Fig. \ref{Fig5}b,e), a positive correlation with properties of the optical paths seems to apply (almost opposite to the previous case). We attribute this pattern to the different distribution of phases in the circuit\cite{Burgwal17}, which was already discussed in Fig. \ref{Fig2}. In particular, we see that the $u_{ij}$  that exhibit a large average deviation from the ideal value are correlated with those with a large number of paths in the mesh. However, here too patterns are more complex and call for more refined, formal analyses to retrace the actual cause. Specifically, it would be interesting to understand why a slightly brighter region appears in the bottom-right corner of the image for $\triangle$ (or a darker region in the center for $\square$).

The above discussion, which involves averages over several unitaries and random initializations, does not say much about specific transformations. Hence, we applied the same analysis to one illustrative example, the Fourier transform, to see whether this is in any way reflected in a single unitary (Fig. \ref{Fig5}c,f). We observe that, despite its symmetry, the sensitivity has the same pattern as the average Haar-random unitary (Fig. \ref{Fig5}b,e).

Overall, this shows that the fidelity is \textit{not} always a suitable metric for high-precision, path-encoded applications. Interestingly, the above patterns survive if we separately plot the two terms in Eq. \ref{eq:estimator_bias}. This means that, for example, multi-photon dynamics would handle noise in a different way, with a marked discrepancy between the most and the least resilient states. This observation will be the object of the next sections.


\subsection{Multi-photon evolution}

\label{sec:noisy_multiphoton}

So far, we discussed how noise affects the matrix elements of $U$, which correspond to single-photon scattering amplitudes. Here we sharpen this analysis by studying how Gaussian noise, applied uniformly on the tunable phases, influences multi-photon transitions. We emphasize that this research question is different from previous analyses in the literature, which looked at global properties of the noisy probability distribution. While this issue may be not too relevant for Boson Sampling, it can be relevant for Scattershot-like applications \cite{Bentivegna15} and for path-encoded implementations of machine learning \cite{Shen17, Flamini19RL, Steinbrecher19, Williamson20, Marquez21, Saggio21, Xu21}, where bias plays a major role. One possible (optimistic) answer we considered was that noise would be almost uniformly distributed on the output, with deviations mildly modulated by properties of $U$. The complementary hypothesis (which turned out to be the case) was that systematic deviations would be found, dependent on $U$ or on the specific transition. In the following we will describe the analysis on a more technical level. 

Let us first introduce some notation. We consider the exponentially large set of input ($\{ v \}$) and output ($\{ w \}$) $N$-photon states (fully indistinguishable photons, without collisions) in an $m$-mode interferometer, and denote it with $\Omega^m_N$. We can associate with each pair ($v$,$w$) a transition probability $p_{vw}$, given by the permanent of a submatrix of $U$  defined by the Fock representation of $v$ and $w$. Looking at the parametric representation of $U$, we have a set of parameters $\{ \xi \} = \{ \psi \} \cup \{ \theta \} $ that describes the action of all phase shifters. For each pair ($v$,$w$), we can also retrieve the subset $\{ \xi \}_{vw} $ that influences $p_{vw}$. Now, we want to understand whether the \textit{signed} deviation of a noisy $\tilde{p}_{vw}$ is on average practically null, or whether there is a systematic bias and, in this case, to what we can attribute it. To this end, we define such deviation as $\delta^{(r)}_{vw} \coloneqq \tilde{p}^{(r)}_{vw} -p_{vw} $, where $r$ enumerates random noisy initializations of $U$, and compute the average $ \delta_{vw} \propto  \sum_r \delta^{(r)}_{vw}   \propto  \big \langle \tilde{p}_{vw} \big \rangle - p_{vw} $ with standard deviation $\sigma_{vw}$.
We then evaluate for all parameters $\xi_k$ the quantity

\begin{equation}
    \Delta_{\xi_k} = \sum_{v,w \in \Omega^m_N} \left( \abs*{\frac{ \delta_{vw} }{\sigma_{vw}} } \frac{\big[ \xi_k \in \{ \xi \}_{vw} \big] }{\abs*{ \{ \xi \}_{vw} }} \right)
    \label{eq:estimator_bias_multiphoton} 
\end{equation}

\noindent where $|\{.\}|$ is the number of elements in the set, $k=1,2,\dots, m(m-1)$ and we used the Iverson bracket

\begin{equation}
    \big[ \xi_k \in \{ \xi \}_{vw} \big] = \begin{dcases}
                    1,                  &  \xi_k \in \{ \xi \}_{vw}    \\
                    0,                  &  \textup{otherwise}
                \end{dcases}.
\end{equation}

\noindent Incidentally, $\delta_{vw} / \sigma_{vw}$ is the inverse of the coefficient of variation of the single deviations that produce  $\delta_{vw} $. The intuition behind this quantity, and more generally about $\Delta_{\xi_k}$, is that we are interested in the significance of the deviation $\delta_{vw}$ from 0 (i.e. how many $\sigma_{vw}$ is $ \delta_{vw} $ far from 0). If there is a relevant systematic bias in the probability distribution, we should be able to observe a preferred (positive or negative) direction for some $p_{vw}$ when we consider all combinations. This real-valued measure of significance is then assigned to each parameter $\xi_k$ that contributed to its magnitude via the Iverson bracket, in proportion to the number of parameters that were involved in that process ($\abs*{ \{ \xi \}_{vw} }^{-1}$). We are now ready to outline how the analysis proceeds.

\begin{figure*}[t]
\includegraphics[width=0.86\textwidth]{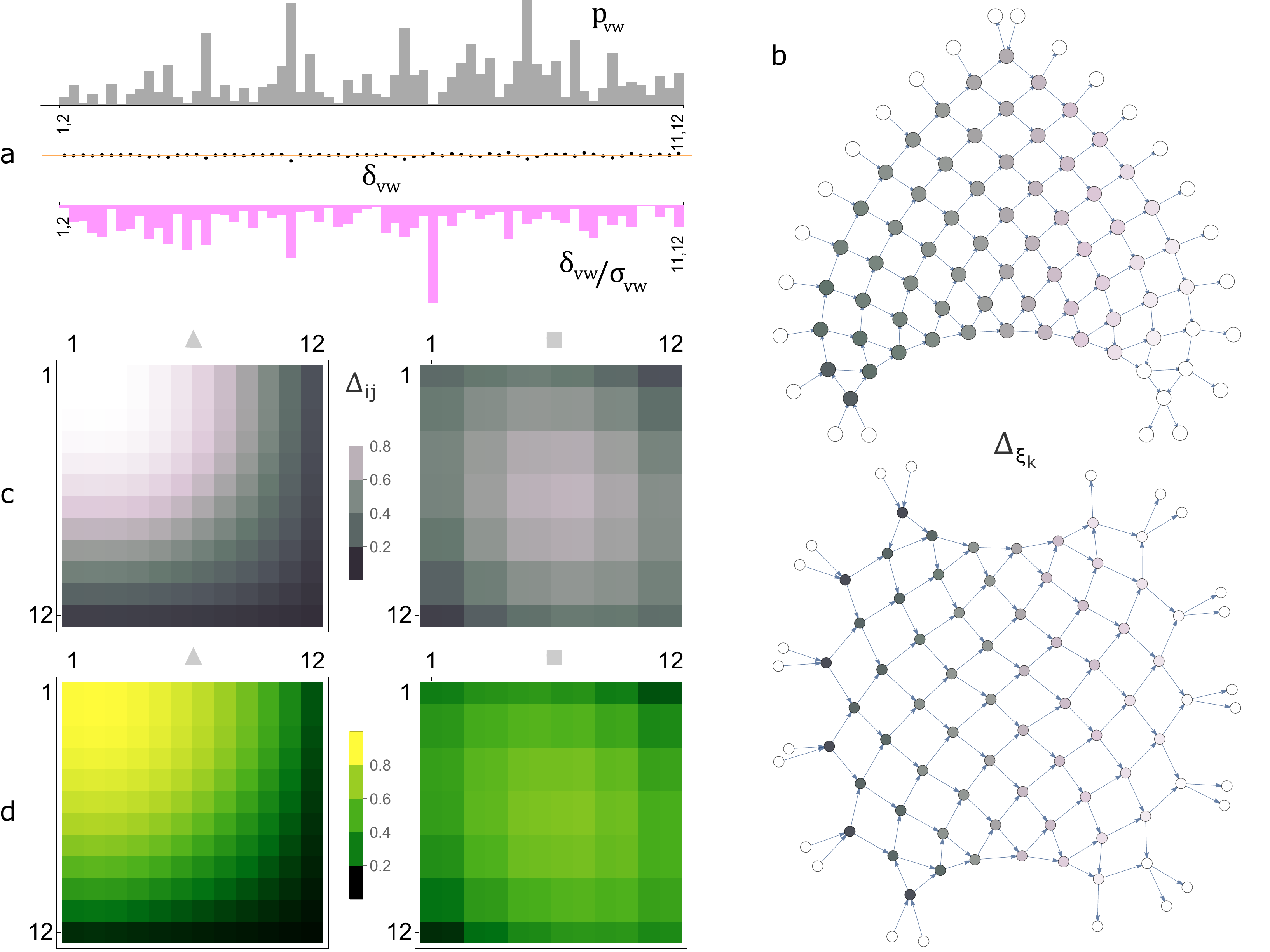}
\caption{\textbf{Sensitivity and multi-photon transitions.} We numerically investigate a systematic deviation from the ideal values of $N$-photon transition probabilities.
a) Three-stage description of the analysis (see Sec. \ref{sec:noisy_multiphoton}). Top panel: (example of) ideal two-photon output (collision-free) probability distribution for a 12-mode Haar-random $U$ from input modes $v=(1,2)$. Middle: for each probability $p_{vw}$ (gray bar), we plot the deviation $ \delta_{vw} = \big \langle \tilde{p}_{vw} \big \rangle - p_{vw} $ (black points) with standard deviation $\sigma_{vw}$, over an orange zero baseline. The take-home message is distilled in the bottom panel. Bottom (mirrored): for each $p_{vw}$, we plot how many $\sigma_{vw}$ separate $ \delta_{vw}$ from 0. Some gray and fuchsia peaks seem to correlate but this is not always true. For $m=12$, $ \delta_{vw} /\sigma_{vw} = 1.49 \pm 0.18$, averaged over all input-output pairs and 200 random initializations of unitaries with Gaussian noise as in Fig. \ref{Fig5}.
b) $\Delta_{\xi_k}$ for all nodes in the $\triangle$ and $\square$ meshes. This pattern is partly due to the fact that cells closer to the output are reached by more input states.
c) $\Delta_{ij}$ normalized to the maximum, for all matrix elements $u_{ij}$. We recover patterns similar to Fig. \ref{Fig3} and \ref{Fig4} for both $\triangle$ (left) and $\square$ (right) meshes.
d) Number of tunable parameters that influence each $u_{ij}$, normalized to the maximum (see also Fig. \ref{FigA2}). This pattern matches the one in panel (c).
}
\label{Fig6}
\end{figure*}

The analysis consists of three stages. (\textit{i}) [\textit{Adding noise}] We pick a Haar-random $U$ and uniformly apply Gaussian noise on all tunable phases (i.e. not on the $u_{ij}$). The analysis was repeated for several $U$ and different sizes $m$, finding a consistent behaviour.  (\textit{ii}) [\textit{From $\tilde{u}_{ij}$ to dynamics}] We numerically generate noisy $N$-photon probability distributions ($N=2,3,4$) and compute $\Delta_{\xi_k} $ using Eq. \ref{eq:estimator_bias_multiphoton} (see Fig. \ref{Fig6}a). (\textit{iii}) [\textit{From dynamics back to $\tilde{u}_{ij}$}] We investigate a possible connection between statistically significant deviations in the probability distribution, tunable phases and $u_{ij}$ (Fig. \ref{Fig6}b,c).
 
An example for this analysis is shown in Fig. \ref{Fig6}a  for $N=2$, $m=12$. For a fixed input state $v'$, we find no obvious correlation between the output states and the magnitude of the bias $\delta_{v'w} \, \sigma_{v'w}^{-1}$: sometimes the bias is large when $ p_{v'w} $ is large, sometimes the converse is true. However, when we consider all input-output combinations as in Eq. \ref{eq:estimator_bias_multiphoton} and ask how bias manifests itself in $U$, we retrieve patterns similar to those in Fig. \ref{Fig3} and \ref{Fig4} - this time from yet another mechanism. To see this, in Fig. \ref{Fig6}c we plot the sum of $\Delta_{\xi_k} $ over all parameters $\xi_k $ that influence $u_{ij}$, namely

\begin{equation}
    \Delta_{ij} = \sum_k  \Delta_{\xi_k} \big[ \xi_k \in \{ \xi \}_{ij} \big] .
\end{equation}

\noindent This connection seems to suggest that a biased error in multi-photon dynamics is correlated with properties of the paths. However, we observe that the pattern in Fig. \ref{Fig6}c matches the distribution of the number of tunable parameters that influence each $u_{ij}$ (Fig. \ref{Fig6}d), providing stronger insights into the underlying error propagation. We emphasize that here we are not simply recombining in $u_{ij}$ noise that was added to each $\xi_k$ that influences $u_{ij}$: in that case, the outcome would be trivial. Instead, we are first mixing the effect of noise in a quantum process (photons do not evolve independently in the circuit), then we are recombining the (inverse of the) coefficient of variation, which is related to a biased error.


\section{Bias in the calibration process}

\label{sec:calibration}

Understanding how a circuit operates is a necessary step for practical applications. With tunable circuits, it is especially necessary to calibrate the voltage-to-phase relationship for each thermo-optic phase shifter in the device. In this manuscript, we will use $\theta \sim \theta_0 + \alpha \, V^2$  and $\psi \sim \psi_0 + \beta \, V^2$ for phase shifters inside and before the MZ, respectively (see Fig. \ref{Fig1}b). During a calibration, the goal is then to estimate ($\theta_0, \alpha, \psi_0, \beta$) for each pair of phase shifters in each MZ. While estimating the transmissivity of directional couplers is also important, here we consider them ideal to isolate the contribution related to the phase.

\begin{figure}[t]
\includegraphics[width=0.99\linewidth]{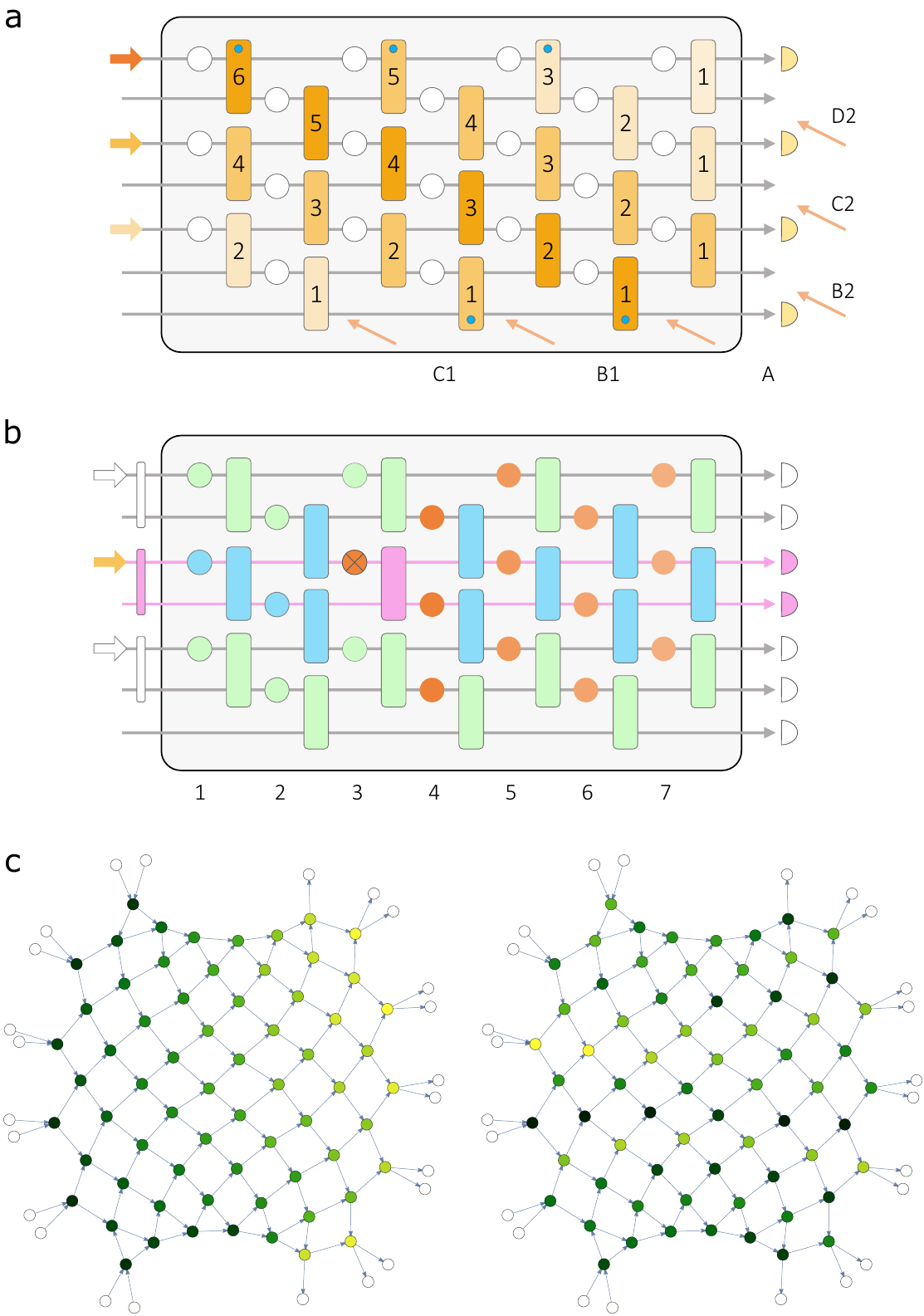}
\caption{\textbf{Calibration and noise.} The voltage-to-phase dependency for the $\square$ mesh can be inferred in two stages (a,b). a) Light is injected in first input mode and measured in the last output, to characterize the diagonal $A$. This step is repeated sequentially for the other diagonals. b) Phase shifters outside the MZs are calibrated by constructing larger MZs consisting of two MZs in the 50:50 state (fuchsia) interleaved with and followed by MZs in the bar state (blue). c) The order chosen by the calibration induces a pattern in the noise distribution. In this example, colors describe the mean deviation of the estimated $\psi_0$ (left) and $\beta$ (right) from the actual values (bright: high; dark: low), averaged over $10^3$ calibrations of Haar-random unitaries.
}
\label{Fig7}
\end{figure}

Bayesian estimation and machine learning are promising approaches to calibrate a device without detailed information on the internal structure (see e.g. Refs. \cite{Cimini21, Fiderer21} and references within).
In the case of $\triangle$ and $\square$ meshes, instead, calibration can be carried out by directing light to each MZ in a specific order.
A strategy to calibrate the $\square$ mesh using knowledge about the structure was briefly sketched in Ref. \cite{Harris17, Taballione19, Taballione20, Arrazola21, Bandyopadhyay21}.  We propose an algorithm in Sec. \ref{sec_app:calibration}, which we use for our discussion on errors (see Fig. \ref{Fig7}). In this scheme, parameters are estimated in two main steps: (\textit{i}) all MZs are probed in sequence (Fig. \ref{Fig7}a), returning an estimate for ($\theta_0, \alpha$); (\textit{ii}) using this information, all MZs are adjusted to either the \textit{bar} or the 50:50 state (Fig. \ref{Fig7}b) to estimate ($\psi_0, \beta$). After this step, additional tricks can be used to improve the estimates. Our main observation, here, is that there is a systematic bias in the uncertainty and accuracy associated with every parameter. This bias is mainly induced by the sequential nature of the calibration, since estimating the latter parameters requires (imperfect) knowledge of several others. In Fig. \ref{Fig7}c, for instance, we show how the average error associated to each MZ correlates with its position in the optical mesh, which in turn influences the sequential procedure of the calibration. This issue can become relevant in large-size circuits.


\section{Discussion} 

Classical and quantum optical technologies provide a promising platform to enable faster and energy-efficient machine learning. Among these technologies, integrated circuits represent a well-established resource with an intuitive operation, good control and advanced miniaturization. 
In this context, while path-encoding is generally a convenient approach to process information, we argue that it is also necessary to counteract the path-dependent noise hidden in the process. This issue manifests itself in an unbalanced control, both in accuracy and precision, over the map between input and output states, which is potentially relevant in tasks such as classification, reinforcement learning or tests on the foundations of physics.  

In this work, we shed light on this aspect by considering single-photon and multi-photon evolutions in triangular and rectangular architectures. We numerically simulate these dynamics in realistic settings, collecting evidence that the impact of noise correlates with properties of the optical paths, as well as with the number of tunable elements they involve. We then present a conjecture for an exact, closed-form expression to compute the number of paths for all input-output combinations and arbitrary size. The validity of this conjecture was verified up to a large circuit size using a graph representation, which we consider of independent interest for several reasons. In fact, we envision the development of techniques to estimate the relevance of nodes in an optical mesh, as done in other fields such as social network analysis. We make a first step also in this direction, reporting on preliminary investigations to identify the most influential regions for error propagation. While some of these tests are inconclusive, we believe that more can be understood by looking carefully with these techniques. This can be especially true for large circuits with arbitrary topology, where special sets of nodes can act as bottlenecks or take on a more influential role for a given application.

\href{https://doi.org/10.5281/zenodo.4924481
}{Code} for all the analyses presented in this manuscript is publicly available online. We conclude with extra remarks on the applicability of this work (see Table \ref{table:applicability}). On the one hand, the positive side is that the imbalance revealed by these analyses seems to be of small magnitude. This means that the above concerns are practically not relevant for most applications today. We leave an absolute estimate of its impact to more specialized studies, which focus on individual technologies and implementations. On the other hand, as the circuits size and the desired accuracy increase, it is reasonable to expect that these concerns will become more relevant in the future. In the highly plausible scenario where optical technologies become a go-to platform for machine learning, similar analyses would also be useful to build trust with the end consumers and funding agencies. We hope this research will help make progress also in this sense.

\begin{table}[h]
    \begin{center} 
        \begin{tabular}{p{0.35\linewidth}p{0.15\linewidth}p{0.15\linewidth}p{0.15\linewidth}}  
        \toprule
                                &  \multicolumn{3}{c}{Term} \\
        \cmidrule(r){2-4}
        Scope                   &  \PBS\centering  Short-       & Medium-         & \PBS\centering  Long- \\
        \midrule
        High-precision tests    & \PBS\centering \cellcolor{gray!10}   & \PBS\centering \cellcolor{gray!10}    & \PBS\centering \cellcolor{gray!10}    \\
        Machine learning        &                       & \PBS\centering \cellcolor{gray!10}    & \PBS\centering \cellcolor{gray!10}    \\
        Others                  &                       &                       & \PBS\centering \cellcolor{gray!10}{\scriptsize Maybe}     \\
        \bottomrule
    \end{tabular}
    \caption{Expected, potential range of applicability of the results presented in this work.}
    \label{table:applicability}
    \end{center}
\end{table}


\section*{Acknowledgements} 

We thank Sebastian Gstir and Robert Keil for their very careful and constructive feedback, which helped improve the manuscript, and Caterina Taballione, Jelmer J. Renema and J\"orn P. Epping for fruitful discussions. We also acknowledge support and stimulating discussions with the Quantum Information \& Computation group and the Photonics group at University of Innsbruck.

The project leading to this application has received funding from the European Union’s Horizon 2020 research and innovation programme under the Marie Sklodowska-Curie grant agreement No 885567 and No 801110, and from the Austrian Federal Ministry of Education, Science and Research (BMBWF).
It reflects only the author's view, the EU Agency is not responsible for any use that may be made of the information it contains. The project has received partial support from the Austrian Science Fund (FWF) through the project SFB BeyondC F71.


\section{Appendix} 

In the following sections, we study three aspects that were only briefly mentioned in the main text. Sec. \ref{sec_app:paths} presents an exact and closed-form expression to compute the number of optical paths inside the canonical meshes. This information should facilitate the development of measures to counteract the influence of biased errors. Sec. \ref{sec_app:centrality} presents considerations on a strategy to estimate the influence of a node on error propagation. Finally, Sec. \ref{sec_app:calibration} completes the discussion started in Sec. \ref{sec:calibration}, presenting a pseudocode to calibrate the $\square$ mesh. This can be useful in practical applications, as well as to understand how a bias originates during the process.


\subsection{Counting optical paths}

\label{sec_app:paths}

Here we present analytical expressions to compute the exact number of paths that light can take from mode $i$ to $j$ in an $m$-mode $\triangle$ and $\square$ optical meshes. We denote this set of paths as $\{\pi_{ij}^m\}$ and its cardinality as $|\{\pi_{ij}^m\}|$. This information is relevant to estimate the susceptibility of each input-output transition to circuit imperfections. We also present analytical expressions for the exact number of paths that connect one input MZ (i.e. a MZ in the first vertical layer) to any MZ inside the $\square$ architecture. This quantity can be used in conjunction with the flow of a MZ (see Eq. \ref{eq:flow}) to assess its criticality in a large mesh. Further work is necessary to connect this result to applications with specific requirements.

\subsubsection{Paths from input to output}

For a fixed mesh, we expect $|\{\pi_{ij}^m\}|$ to depend on the horizontal layers to which $i$ and $j$ belong (modes in the same input/output MZ have the same number of paths) and on the parity of $m$, since $\square$ has horizontal symmetry when $m$ is odd. Below we present an analytic expression that indeed fulfills these observations.

Given its formulation, we approach the problem using combinatorics. We find \textit{by inspection} that a solution can be expressed using Catalan’s trapezoids\cite{Reuveni14} $C_s (a,b)$, a generalization of the Catalan’s triangle (in turn generalizing the Catalan numbers, defined using binomial coefficients). A closed-form expression for $C_s (a,b)$ is

\begin{equation}
    C_s (a,b) = \begin{dcases}
                    \binom{a+b}{b},                     &  0 \leq b < s \\
                    \binom{a+b}{b}-\binom{a+b}{b-s},    &  s \leq b \leq s+a-1 \\
                    0,                                  &  b > s+a-1
                \end{dcases}
    \label{eq:catalantrap}
\end{equation}

\noindent where $s=1,2,3,\dots$, $a=0,1,2,\dots$ and $b=0,1,2,\dots$. We find that, by introducing the compact notation

\begin{equation}
    \begin{split}
        \sigma^m_{ij} &\coloneqq \frac{1}{2} \, \textup{sgn} \Big(\frac{1}{2} + i + j - m  \Big)  \\
        T^m_{ij} &\coloneqq -i + j + \frac{1}{2}(-1)^i + \frac{1}{2} (-1)^{m+j}
    \end{split}
\end{equation}

\noindent the combination $(S^m_{ij}, A^m_{ij}, B^m_{ij})_\square$, with

\begin{equation} \label{eq:snk}
    \begin{split}
        S^m_{ij} &\coloneqq \frac{m + 2}{2}  + \sigma^m_{ij} \Big(m - 2i + (-1)^i +1 \Big) \\
        A^m_{ij} &\coloneqq \frac{m-1}{2} - \sigma^m_{ij} \, T^m_{ij}   \\
        B^m_{ij} &\coloneqq \frac{m-1}{2} + \sigma^m_{ij} \, T^m_{ij} ,
    \end{split}
\end{equation}

\noindent indeed provides the answer to our counting problem 

\begin{equation}
        |\{\pi_{ij}^m\}|^{\square} = C_{S^m_{ij}} (A^m_{ij}, B^m_{ij}) .
    \label{eq:pathsC}
\end{equation}

\noindent Recalling that Eq. \ref{eq:catalantrap} is a piecewise function over three regions, we observe that the last condition is always satisfied for all values given by Eq. \ref{eq:snk}, due to the constraints of the problem ($1 \leq i,j \leq m$). From a physical perspective this makes sense, since $|\{\pi_{ij}^m\}|^{\square} $ cannot be 0.

\begin{figure}[t]
\includegraphics[width=0.95\linewidth]{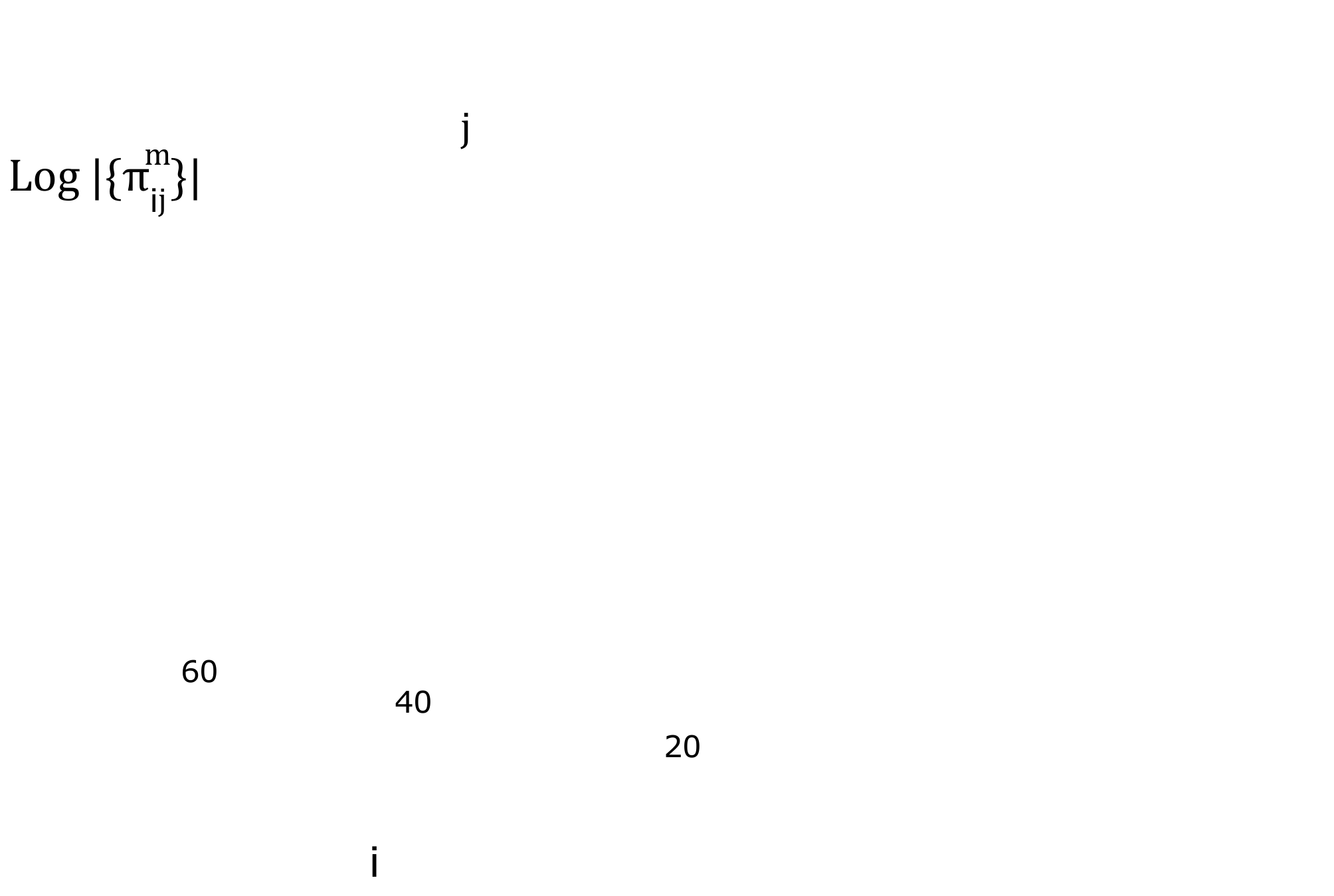}
\caption{\textbf{$\square$ and number of paths.} Logarithm of the number of paths $|\{\pi_{ij}^m\}|^{\square}$ that connect mode $i$ to mode $j$ in the $\square$ mesh using Eq. \ref{eq:pathsC}. Three surfaces are shown for a different size of the mesh ($m=20, 40, 60$). The same plot for $\triangle$ would exhibit peaks on one of the corners, as in Fig. \ref{Fig3}.
}
\label{FigA1}
\end{figure}

We do not know why Eq. \ref{eq:pathsC} is true, but we numerically verified its correctness for all $i,j \in [ 1,m ]$ and $m \in [ 1,27 ]$. We leave the pleasure of the proof to the curious and more skilled reader. For circuits with larger size ($m \sim30$) it becomes computationally challenging to list all paths on a workstation with 36 cores and 128 GB of RAM. 

By a similar counting argument, choosing

\begin{equation}
    (S_{i}, A^m_{i},B^m_{j})_\triangle=(i,m-i,m-j)
\end{equation}

\noindent we also find the expression for the $\triangle$ mesh

\begin{equation}
    \begin{split}
          |\{\pi_{ij}^m\}|^{\triangle} = \binom{2m -i-j}{m-j}-\binom{2m -i-j}{m}
    \end{split}
    \label{eq:pathsR}
\end{equation}

\noindent Notice that, due to the symmetry rule for binomial coefficients, Eq. \ref{eq:pathsR} is symmetric in $i$ and $j$, as expected.

\begin{figure}[b]
\includegraphics[width=0.975\linewidth]{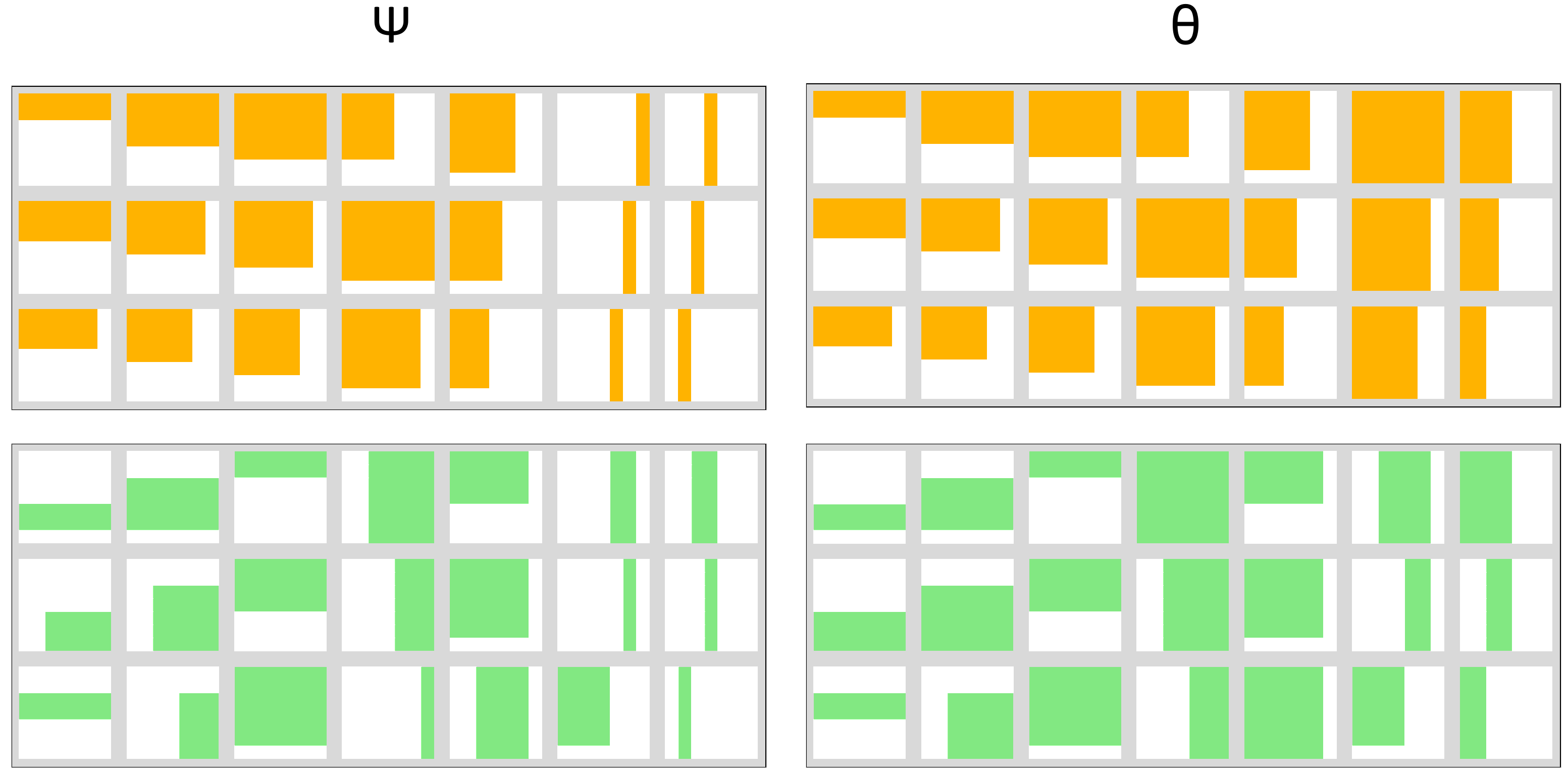}
\caption{\textbf{Parameters distribution.} Each matrix element of $U$ depends on different sets of phase shifters ($\psi$) and beamsplitters ($\theta$). As an example, here we consider $m=7$ in the $\triangle$ (orange) and $\square$ (green) meshes. In each panel, each square $k$ represents a $7 \times 7$ matrix with matrix elements colored if they depend on $\theta_k$ or $\psi_k$ ($k=1,2,\dots,21$).
}
\label{FigA2}
\end{figure}

We emphasize that $|\{\pi_{ij}^m\}|^{\square}$ and $|\{\pi_{ij}^m\}|^{\triangle}$ provide the exact number of paths between modes ($i$, $j$). Using these quantities, we can quantify the asymmetry between any single-photon transitions. We can also combine this information to study multi-photon processes with distinguishable photons (since they evolve independently), while correlations may be found in dynamics that involve quantum interference. Indeed, we recall that each $u_{ij}$ depends only on a subset of tunable parameters, related to the waveguide connectivity in the circuit (see Fig. \ref{FigA2}).

We can also use Eq. \ref{eq:pathsC} and \ref{eq:pathsR} to quantify the asymmetry between architectures w.r.t the number of paths that connect two modes. This information can be useful since $|\{\pi_{ij}^m\}|$ correlates with the resilience to noise, which can be very different in the two cases. For instance, for $m$ large and $i=j=1$, using the Stirling approximation for the binomial coefficient we obtain the scaling

\begin{equation}
    \frac{|\{\pi_{11}^m\}|^{\triangle} }{|\{\pi_{11}^m\}|^{\square} } \sim \frac{ 2^{m-\frac{3}{2}} }{ m }
    \label{eq:scaling11}
\end{equation}

\noindent which shows an exponential separation between the two architectures (see also Fig. \ref{Fig3}). Similarly, when $i=1$ and $j=\floor*{\frac{m}{2}}$ we can use

\begin{equation}
    \binom{a+b}{b} \sim \sqrt{\frac{a}{2 \pi (a-b)}} \, \frac{a^a}{b^b (a-b)^{a-b}},
\end{equation}

\noindent which works when both $a$ and $b$ are large, and $\Gamma(x) \sim \sqrt{2\pi} \, x^{x-\frac{1}{2}} \, e^{-x}$, to get

\begin{equation}
    \frac{|\{\pi_{1\frac{m}{2}}^m\}|^{\triangle} }{|\{\pi_{1\frac{m}{2}}^m\}|^{\square} } \sim  e^{w m -1 },
    \label{eq:scaling1m2}
\end{equation}

\noindent with $w=\log{3/8} + 5/4 \log{3} \sim 0.39$. We tested the approximations in Eq. \ref{eq:scaling11} and \ref{eq:scaling1m2} in the regime $ 30 < m < 300 $, finding a very good agreement with the exact value given by Eq. \ref{eq:pathsC} and \ref{eq:pathsR}. Together, they provide quantitative understanding of the asymmetry between the two architectures. This knowledge can guide the choice of the most appropriate platform for a given application that needs to fulfill specific physical constraints, as well as to correct for this inherent imbalance.


\subsubsection{Paths from a Mach-Zehnder to input and output}

Here we discuss how to compute the number of paths that connect any MZ in the $\square$ architecture (located in layer $l$, $q$-th position from top) with the input and the output layers of the circuit. In the representation based on a directed graph, we would count in how many ways we can reach a node from the input ($|\{\pi_q^{l,m}\}|_{IN}$), as well as in how many ways $n$ can reach the output ($|\{\pi_q^{l,m}\}|_{OUT}$). Intuitively, this corresponds to the number of paths that lay in the light cone of a MZ (summarized by $I_n$ and $O_n$ in Sec. \ref{sec:graphs}). This information can help evaluate to what extent a MZ affects scattershot-like applications \cite{Bentivegna15}, where light is injected and measured in all input and output modes. We will return to this point in Sec. \ref{sec_app:centrality}.

We tackle this task by considering the number of paths ($|\{\pi_{q'q}^{l,m}\}|$) that connect a MZ in the first layer ($q'$-th from top) with any MZ in the circuit ($l\geq 1$, $q$-th from top). Notice that each MZ in the first layer can be reached by two input modes, so a factor of $2$ may be introduced to describe the problem in terms of modes. In both cases, we consider again the Catalan's trapezoid in Eq. \ref{eq:catalantrap} and look for a combination $(S^{l,m}_{q'q}, A^{l,m}_{q'q}, B^{l,m}_{q'q})_\square^G$ that fits the problem. Here, the superscript $G$ serves to remember that we are describing the problem in terms of nodes in a graph. After some experimentation, we find by inspection that the following expressions

\begin{equation} \label{eq:snk_nodes}
    \begin{split}
        S^{l,m}_{q'q} &\coloneqq \frac{m + 2}{2}  + \sigma^m_{q'q} \Big(m - 2q' + (-1)^{q'} - (-1)^l \Big) \\
        A^{l,m}_{q'q} &\coloneqq \frac{l-1}{2} - \sigma^m_{q'q} \, \Big( T^m_{q'q} - (l-1) \textup{ mod } 2 \Big) \\
        B^{l,m}_{q'q} &\coloneqq \frac{l-1}{2} + \sigma^m_{q'q} \, \Big( T^m_{q'q} - (l-1) \textup{ mod } 2 \Big)
    \end{split}
\end{equation}

\noindent solve the problem for circuits with odd size $m$, i.e.

\begin{equation}
        |\{\pi_{q'q}^{l,m}\}| = C_{S^{l,m}_{q'q}} (A^{l,m}_{q'q}, B^{l,m}_{q'q}).
    \label{eq:pathsMZ}
\end{equation}

\noindent Similar considerations lead to an analogous of Eq. \ref{eq:pathsMZ} for even $m$ and for the $\triangle$ architecture.

Indeed, by using Eq. \ref{eq:catalantrap} it is not hard to see that for $l=1$ we have $|\{\pi_q^{1,m}\}|_{IN}=0$, since  $A^{1m}_{q'q} = -B^{1m}_{q'q} = \sigma^m_{q'q} \, T^m_{q'q} $ (no path connects two MZs in the same layer). In addition, when  $l=m=2r+1$ ($r=0,1,2,\dots$) we retrieve the same expression in Eq. \ref{eq:pathsC}, as desired. As mentioned, the only difference is that Eq. \ref{eq:pathsC} refers to input-output modes while here we refer to MZs. To match Eq. \ref{eq:pathsC} and Eq. \ref{eq:snk_nodes} it is sufficient to observe that $q=\floor*{\frac{i+1}{2}}=1,2,\dots,\frac{m-1}{2}$ and $q'=\floor*{\frac{j+1}{2}}=1,2,\dots,\frac{m-1}{2}$.

By using Eq. \ref{eq:pathsMZ} (odd size) and the symmetry of the problem, we can now evaluate

\begin{equation} \label{eq:pathsMZtotalINOUT}
    \begin{split}
        |\{\pi_q^{l,m}\}|_{IN} &= \sum_{q' =1}^{\frac{m-1}{2}} C_{S^{l,m}_{q'q}} (A^{l,m}_{q'q}, B^{l,m}_{q'q}), \\
        |\{\pi_q^{l,m}\}|_{OUT} &= \sum_{q' =1}^{\frac{m-1}{2}} C_{S^{m-l,m}_{q'q}} (A^{m-l,m}_{q'q}, B^{m-l,m}_{q'q}),
    \end{split}
\end{equation}

\noindent which provides a baseline to study the centrality of a node. This task will be the focus of the next section.


\subsection{Centrality measures in an optical mesh}

\label{sec_app:centrality}

\begin{figure*}[t]
\includegraphics[width=0.99\textwidth]{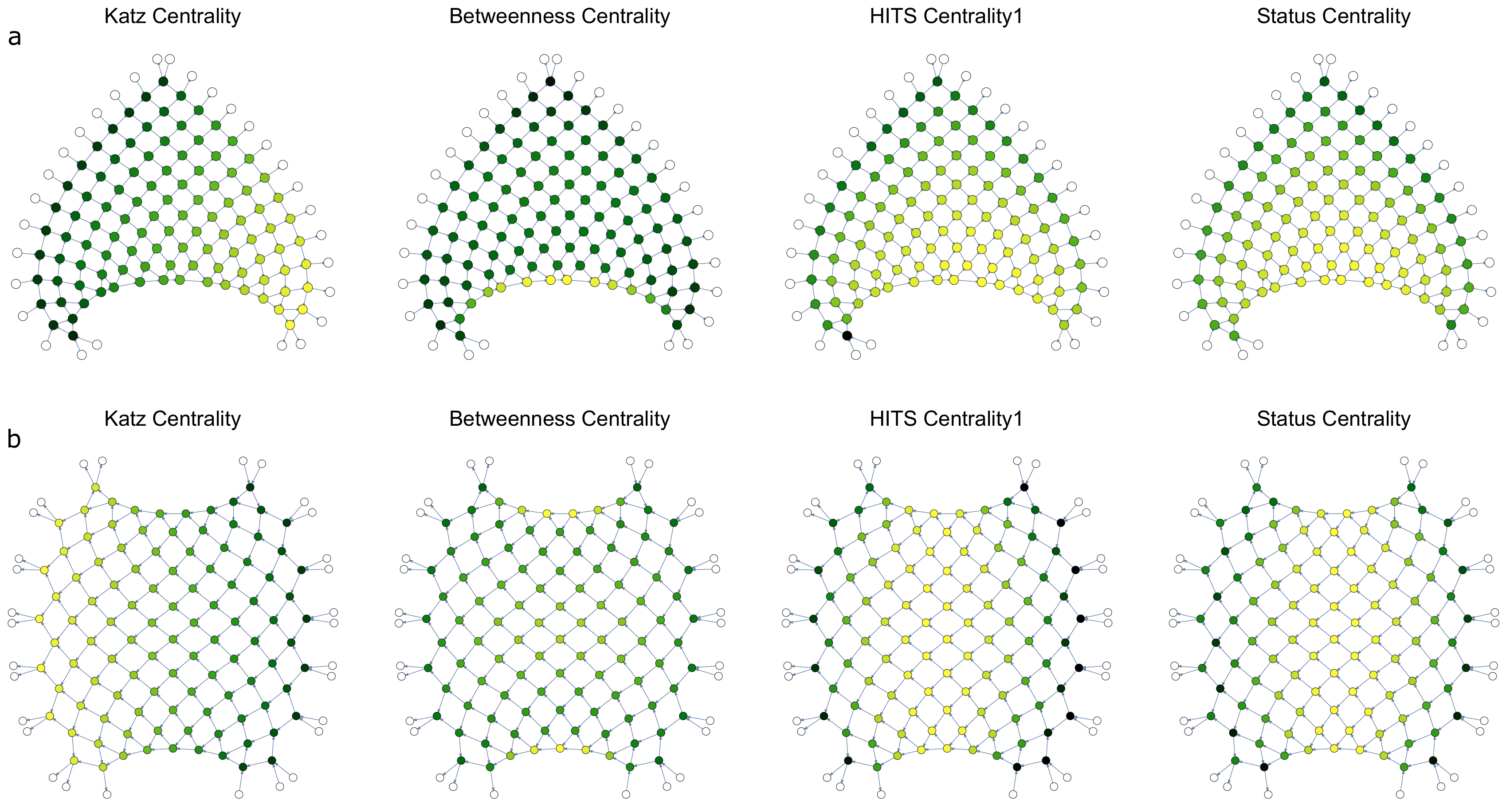}
\caption{\textbf{Centrality of a Mach-Zehnder.} Centrality measures of each node of the optical mesh are displayed with a color scheme, from low (green) to high (yellow), for the $\triangle$ (a) and $\square$ (b) meshes. Absolute values of the centrality are not relevant.
}
\label{FigA3}
\end{figure*}

In this section, we turn our attention to the impact of \textit{individual} MZs in realistic implementations. The objective is to identify the most important nodes within the optical meshes, according to a suitable real-valued measure of importance. To this end, motivated by the analogy between light propagation in a noisy circuit (where each MZ can introduce noise into the evolution) and information spreading, here we will consider tools from graph theory and network analysis, where this notion of importance is generally called \textit{centrality}. The research question we ask is whether network analysis can help understand what noisy MZs are more likely to influence the performance of a circuit, without any information on the underlying quantum process. In other words, is it possible to infer the influence of a MZ by just looking at the network topology?

Centrality measures rank nodes according to their flow (for different types of flow) or to their involvement in the cohesiveness of the network. They are only good within the application they were designed for, and they are likely to disagree otherwise. These rankings provide the order of the most important nodes (although not the relative importance), while little can be said for the majority of the others. This limitation has led to the development of measures to assess the influence of all nodes in a network, e.g. the \textit{accessibility} (how accessible the network is from a node using random walks) and the \textit{expected force} of infection. This metric is more related to the flow discussed in Sec. \ref{sec_app:paths} and in Eq. \ref{eq:flow}.
Another aspect of centrality measures is that they operate from a node-centric perspective, i.e. they don't consider the synergy between neighbor nodes (which plays a major role in disease and information spreading). In this case, game-theoretic approaches are usually more effective. Centrality and node influence can be correlated but they are in general distinct. 

Given these considerations, we computed various measures of centrality for each MZ \cite{Mathematica}: closeness centrality, betweenness $-$, degree $-$,  eigenvector $-$,  Katz $-$,  PageRank $-$,  HITS $-$,  radiality $-$,  status $-$, edge betweenness $-$, and LinkRank $-$. Results for some of them - those with the most interesting behaviour - are shown in Fig. \ref{FigA3}. Contrary to our expectation, we did not find a strong correlation with the reported results.

We also carried out additional numerical simulations to single out the most influential MZs, by adding realistic levels of Gaussian noise on each phase shifter one by one (assuming all the other elements are ideal or nearly ideal). In all these cases, a Monte Carlo simulation using the Fidelity as figure of merit did not reveal any consistent pattern of resilience: the reduction in the Fidelity was too small, comparable to the average fluctuation. We also simulated an imperfect implementation where the error on each phase was proportional to the magnitude: in this case, we indeed recovered the expected pattern due to the phase distribution for Haar-random unitaries \cite{Burgwal17}. We believe that more informative analyses should be tailored to specific architectures, to circumvent the bias induced by the Haar measure.


\subsection{Calibration}

\label{sec_app:calibration}

In this section we outline an algorithm to calibrate $\square$ meshes. While the idea was already sketched in previous works\cite{Harris17, Taballione19, Taballione20, Arrazola21, Bandyopadhyay21}, to the best of our knowledge no detailed procedure was proposed that is amenable to immediate reuse and improvement. Understanding this operation is also useful to evaluate the results in Fig. \ref{Fig7}, and gauge how different approaches may induce different errors.

In principle, calibration could also be seen as an optimization problem, where one looks for the set of parameters that best reproduces all measurements, according to a suitable figure of merit. To get a rough understanding of the feasibility of this approach, we attempted a calibration with a simple, unpretentious least-squares optimization (LSO). Specifically, we (\textit{i}) considered a $\square$ circuit of increasing size $m$, (\textit{ii}) simulated single-photon ($P_1$) and two-photon ($P_2$) output probability distributions for a set of random parameters $\{\xi\}$, and (\textit{iii}) compared $P_1$, $P_2$ with some $P'_1$ and $P'_2$ obtained by adding Gaussian noise on both transmissivities and phases. Before this preliminary test, our guess was that, for sufficiently small $m$ and low noise, a LSO would return a good estimate of the noiseless parameters. Contrary to our expectation, we found that, even though LSO does a very good job in minimizing the residuals, it returns parameters that can be quite different from the original ones. In addition, both the quality of the result and the required evaluation time become rapidly unfavorable for circuit sizes as low as $m \sim 7$ (the number of parameters scales quadratically in $m$). While we emphasize that one cannot draw any conclusive consideration from this simple test, we gain evidence that (a) specialized strategies are likely to perform significantly better, and (b) simple LSO-like calibrations can fail due to the large number of parameter combinations that approximate the desired output. Importantly, (c) using the output of a LSO to study specific input-output scattering processes may result in large errors, since $\{\xi\}$ was estimated from a global optimization problem.

The proposed approach consists of two parts (see Algorithm 1). (\textit{i}) The first two loops serve to characterize the phase shifters in each MZ, by sequentially focusing on the elements below and above the main diagonal of the circuit, respectively. Specifically, each iteration selects a path in the circuit and calls one subroutine (Algorithm 2) that characterizes the phase shifters inside the MZs by LSO. This step reveals what voltages correspond to the \textit{bar} ($V^{-}$; $\textup{MZ}_{-}$), \textit{cross} ($V^{\times}$; $\textup{MZ}_{\times}$) and balanced ($V^{50}$) states, which allow to guide light to specific regions of the circuit. Some of these steps can commute and can be parallelized for a faster execution on large circuits. (\textit{ii})  The last two loops in Algorithm 1 provide information on the phase shifter before each MZ. First, the linear dependency on the applied voltage is characterized (yielding the $\beta$ coefficients). Once these are known, we can remove the voltage and focus on the offsets $\psi_0$, which completes the calibration.


\begin{algorithm}[H]
   \begin{algorithmic}[1]
    \caption{Calibration}
    \Require{Graph $G$ for a rectangular\cite{Clements16} architecture of size $m$. } 
    \Ensure{An estimate of the voltage-to-phase relationship for all tunable phase shifters.}
    \Statex
    \State $N_{\textup{MZ}} \gets \frac{m(m-1)}{2}$
    \State MZpath, $\textup{MZ}_{-}$  $\gets$ Empty lists
    \State $ \textup{Model}_\psi \; \gets \frac{1}{2}+ s \, \cos ( 2 \psi_0 + 2 \beta V_\psi ) + t \, \sin ( \psi_0 + 2 \beta V_\psi )$
    \Statex
    \For {$d \gets 1$ to $\floor{\frac{m}{2}}$}
        \State $\textup{MZ}_{-}$ $\gets$ $\textup{MZ}_{-}$ $\cup$ MZpath
        \State MZpath $\gets$ Shortest path in $G$ from $2 d-1$ to $N_{\textup{MZ}}$
        \State FitPath($2 d - 1, m$)
    \EndFor
    \Statex
    \For {$d \gets 1$ to $\floor{\frac{m-1}{2}}$}
        \State $\textup{MZ}_{-}$ $\gets$ $\textup{MZ}_{-}$ $\cup$ MZpath
        \State MZpath $\gets$ Shortest path in $G$ from $1$ to $N_{\textup{MZ}} - 2 d + 1$
        \State FitPath($1, m - 2 d + 1$)
    \EndFor
    \Statex
    \For {vertical layer $l_v$ in $G$}
        \ForAll{$k \in$ 1 to $N_{\textup{MZ}}$}
            \State $ V_k^{(\theta)} \gets \left\{
                        \begin{array}{@{}ll@{}}
                            V^{50}_k,       &  k \in l_v \\
                            V^{-}_k,        &  \textup{otherwise} \\
                        \end{array}\right. $
            \State $ V_k^{(\psi)} \gets \left\{
                        \begin{array}{@{}ll@{}}
                            V_\psi,  &  k \in l_v \\
                            0,          &  \textup{otherwise} \\
                        \end{array}\right. $            
        \EndFor
        \For {horizontal layer $l_h$ in $G$}
            \State Inject light in $l_h$ and measure output distribution.
            \State $\beta^{(k)} \gets$ Scan $V_\psi$ and fit data using $\textup{Model}_\psi$.
        \EndFor
    \EndFor
    \Statex
    \For {vertical layer $l_v$ in $G$}
        \ForAll{$k \in$ 1 to $N_{\textup{MZ}}$}
            \State $ V_k^{(\theta)} \gets \left\{
                        \begin{array}{@{}ll@{}}
                            V^{50}_k,       &  k \in l_v \\
                            V^{-}_k,        &  \textup{otherwise} \\
                        \end{array}\right. $
            \State $ V_k^{(\psi)} \gets 0$            
        \EndFor
        \For {horizontal layer $l_h$ in $G$}
            \State Inject light in $l_h$ and measure output distribution.
            \State $\psi_0^{(k)} \gets$ Approximate distribution using $\textup{Model}_\psi$.
        \EndFor
    \EndFor
    \end{algorithmic}
\end{algorithm}

To test the algorithm, we numerically simulated a full calibration for (several random instances of) noisy settings and circuits with size up to $m=15$ and realistic levels of Gaussian noise. For the sake of convenience, we make three  assumptions: (a) we assume $0 \leq \psi_0^{(k)} \leq \frac{\pi}{4}$ (to remove some ambiguity in the determination of the offset); (b) we assume that it is possible to inject light from pairs of input modes with a known phase relationship (which we set to zero w.l.o.g.); (c) directional couplers are ideal (an imperfect fabrication can be included in Algorithm 1). Condition (b) is illustrated in Fig. \ref{Fig7} by means of an extra beamsplitter before the circuit, which splits the input light into two balanced paths that recombine in one of the MZs, thus forming a larger MZ.


\begin{algorithm}[H]
  \caption{FitPath \textemdash	 {\small Pseudocode for the subroutine called by Algorithm 1 to calibrate the diagonals.} }
   \begin{algorithmic}[1]
    \Require{Input mode $I$; Output mode $O$} 
    \Ensure{$(\theta_0^{(k)}, \alpha^{(k)}, V^{-}_k$, $V^{\times}_k, V^{50}_k)$ for each MZ $k$.}
    \Statex
    \State $\textup{MZ}^{\times}$ $\gets$ Empty list
    \State $ \textup{Model}_\theta \; \gets \frac{1}{2} \left(1+ \cos ( 2 \theta_0 + 2 \alpha V_\theta ) \right)$
    \State FitMZ($b$) $\gets$ 1. Inject light in $I$; 2. Measure light from $O$; 3. Scan voltage $V_\theta$ in MZ $b$; 4. Fit data using $\textup{Model}_\theta$.
    \Statex
    \For{$b \in$ Reverse(MZpath)}
        \State $\textup{MZ}^{\times}$ $\gets$ $\textup{MZ}^{\times}$ $\cup$ $\{b\}$
        \ForAll{$k \in$ 1 to $N_{\textup{MZ}}$}
            \State $ V_k^{(\theta)} \gets \left\{
                        \begin{array}{@{}ll@{}}
                            0,              &  k \notin \textup{MZ}^{-} \cup \textup{MZ}_{\times} \\
                            V^{-}_k,        &  k \in \textup{MZ}^{-} \wedge  k \neq b \\
                            V^{\times}_k,   &  k \in \textup{MZ}^{\times} \wedge  k \neq b
                        \end{array}\right. $
            \State $ V^{(\psi)}_k \gets  0 $
        \EndFor
        \State FitMZ($b$)
        \State Evaluate $V^{-}_b$, $V^{\times}_b$, $V^{50}_b$ by inverting $\textup{Model}_\theta$.
        \State Append values $V^{-}_b$, $V^{\times}_b$, $V^{50}_b$ to lists $V^{-}$, $V^{\times}$, $V^{50}$.
    \EndFor
   \end{algorithmic}
\end{algorithm}


In our numerical tests, Algorithm 1 and 2 retrieved very good estimates of the actual (unknown) voltage-to-phase relationship, for different $m$ and levels of noise, with a fully automated procedure. We also observe how the quality of the characterized unitary (with no voltage) decreases for larger $m$, especially since errors propagate through the sequential calibration. These errors can be mitigated in various ways: (\textit{i}) by repeating the procedure multiple times, choosing a different order where possible, and averaging the estimates; (\textit{ii}) using optimization algorithms, e.g. particle swarm optimization, starting from the parameters given by the calibration (the Fidelity can be used a simple loss function);
(\textit{iii})  exchanging the input and the output and averaging the estimates.



\end{document}